\documentclass[
reprint,
superscriptaddress,
showkeys,
showpacs,
nofootinbib,
aps,
prl
]{revtex4-1}

\usepackage[utf8]{inputenc}
\usepackage[english]{babel}
\usepackage{amsmath}
\usepackage{amsfonts}
\usepackage{amssymb}
\usepackage{graphicx}

\usepackage{color}

\usepackage{upgreek}

\newcommand{\bfA}{{\mathbf{A}}}

\newcommand{\bfB}{{\mathbf{B}}}

\newcommand{\bfD}{{\mathbf{D}}}
\newcommand{\bfE}{{\mathbf{E}}}

\newcommand{\bfH}{{\mathbf{H}}}

\newcommand{\bfp}{{\mathbf{p}}}
\newcommand{\bfr}{{\mathbf{r}}}

\newcommand{\bfv}{{\mathbf{v}}}

\newcommand{\rmd}{{\mathrm d}}

\newcommand{\rme}{{\mathrm e}}

\newcommand{\rmg}{{\mathrm g}}
\newcommand{\rmi}{{\mathrm i}}

\newcommand{\rmr}{{\mathrm r}}

\newcommand{\bcE}{{\boldsymbol{\mathcal E}}}

\newcommand{\bcH}{{\boldsymbol{\mathcal H}}}
\newcommand{\cI}{{\mathcal I}}

\newcommand{\cV}{{\mathcal V}}

\newcommand{\bcA}{{\boldsymbol{\mathcal A}}}
\newcommand{\bfq}{{\mathbf{q}}}

\newcommand{\rot}{{\mathrm{rot}} \, } 
\newcommand{\derivep}[2]{ \frac{\partial #1}{\partial #2} }
\newcommand{\deriv}[2]{ \frac{\mathrm{d} #1}{\mathrm{d} #2} }

\usepackage{graphicx}
\makeatletter
\let\saved@includegraphics\includegraphics
\AtBeginDocument{\let\includegraphics\saved@includegraphics}
\renewenvironment*{figure}{\@float{figure}}{\end@float}
\makeatother

\begin{document}

\title{Universality of the Abraham-Minkowski dilemma for photon momenta beyond dielectric materials}

\author{Damien~F.~G.~Minenna}%
 \email[Corresponding author.\\]{damien.minenna@univ-amu.fr}
\affiliation{%
Centre National d'{\'E}tudes Spatiales, FR-31401 Toulouse cedex 9, France
}%
\affiliation{%
Aix-Marseille Universit{\'e}, CNRS, PIIM, UMR 7345, FR-13397 Marseille, France
}%
\affiliation{%
Thales, FR-78140 V{\'e}lizy, France
}%
\author{Yves~Elskens}%
\affiliation{%
Aix-Marseille Universit{\'e}, CNRS, PIIM, UMR 7345, FR-13397 Marseille, France
}%
\author{Fabrice~Doveil}%
\affiliation{%
Aix-Marseille Universit{\'e}, CNRS, PIIM, UMR 7345, FR-13397 Marseille, France
}
\author{Fr{\'e}d{\'e}ric~Andr{\'e}}%
\affiliation{%
Thales, FR-78140 V{\'e}lizy, France
}%

\date{\today{} \copyright The authors}

\begin{abstract}
Whenever light is slowed down, for any cause, two different formulas give its momentum.
For dielectrics, the coexistence of those momenta was the heart of the century-old Abraham-Minkowski dilemma, recently resolved.
We demonstrate that this framework extends to momentum exchange in wave-particle interaction; in particular to Langmuir waves for Landau damping and to vacuum waveguides of electron tubes (metallic slow-wave structures).
Focussing on the latter, we show that the dilemma resolution is not limited to discriminating between kinematic and canonical momenta but also involves a non-negligible momentum flux from Maxwell's electromagnetic stress.
The existence of two momenta in materials, plasmas, and waveguides, for which light velocity modification has entirely different origin,
points to the universality of the Abraham-Minkowski dilemma.
\end{abstract}

    
\maketitle

The question ``what is the momentum of light propagating through glass?'' seems simple enough to appear on a high-school quiz \cite{cho10,leo06}. 
Nevertheless, it took physicists more than a hundred years to answer it in a vigorous debate 
confronting two rival theories and called the Abraham-Minkowski controversy.
For Minkowski \cite{min08}, respectively Abraham \cite{abr09}, the momentum carried by light is proportional ($p_{\rm Mink} = n \hbar \omega / c$), respectively inversely proportional ($p_{\rm Abra} = \hbar \omega / ( n c)$), to the refractive index $n$ of dielectric and magnetic materials (see Table~\ref{t:table1}).
The problematic is the same for classical fields or photons.
Despite many theoretical and experimental works \cite{jon54,jon78,bre79,pfe07}, 
it took a century until recent contributions \cite{hin09,bar10,bar10b} proposed a resolution of the controversy, arguing that \textit{both} theories are correct.
While this resolution is largely accepted, the dilemma is still actively debated in materials \cite{sil17,bre18}.

Mathematically speaking, the notion of momentum admits two definitions.
The linear (kinematic) momentum of a particle ($i$) is its mass times its velocity ($\bfp_{{\rm kin},i}  = m \dot{\bfq_i}$, linked to forces), while the conjugate (canonical, in Hamiltonian formalism) momentum is the derivative of the Lagrangian ($\bfp_{{\rm can},i} = \partial_{\dot{\bfq}_i} L$, linked to action).
In newtonian mechanics, there is no difference between both momenta, 
but this is no more the case in quantum physics or in electrodynamics.
In that respect, the canonical momentum \cite{jac99} of a charged particle ($i$) immersed in a magnetic field is $\bfp_{{\rm can},i} = \bfp_{{\rm kin},i} + e \bfA(\bfq_i)$, 
with its electric charge $e$, and the vector potential $\bfA$ such that $\bfB = \rot \bfA$.
The discrepancy between both momenta comes from the involvement of the field. 
Therefore, we should also expect discrepancies between momenta of light.  
This is precisely where the controversy originates: 
the canonical momentum of light is Minkowski's expression $\bfp_{\rm Mink}$  
while Abraham's expression $\bfp_{\rm Abra}$ is kinematic. 
According to Refs.~\cite{bar10,bar10b}, in dielectric and magnetic media, both momenta are linked via the total momentum $\bfp^{\rm matter}_{{\rm can}} + \bfp_{{\rm Mink}} = \bfp^{\rm matter}_{{\rm kin}}  +  \bfp_{\rm Abra}$, with  the canonical $\bfp^{\rm matter}_{{\rm can}}$ and kinematic $\bfp^{\rm matter}_{{\rm kin}}$ momenta of matter.

\begin{table}
\centering
\caption{\label{t:table1} Established expressions \cite{pfe07,mil10} for the momentum of light and its density, in dielectric and magnetic materials.
A single photon has wavenumber $k = \omega / v_\phi = n \omega / c$, with angular frequency $\omega$, phase velocity $v_{\phi}$, and refractive index $n=c/v_\phi$. 
Classical electrodynamics of media calls on the electric field $\bfE$, the magnetic field $\bfH$, the electric displacement $\bfD = \epsilon \bfE$, the magnetic induction $\bfB = \mu \bfH$, the permittivity $\epsilon = \epsilon_\rmr \epsilon_0$, the magnetic permeability $\mu = \mu_\rmr \mu_0$, and the speed of light in vacuum $c = (\epsilon_0 \mu_0)^{-1/2}$.
The group index of a dispersive medium is $n_\rmg = c / v_\rmg$, with group velocity $v_\rmg = \partial_k \omega$.
Since the group index is $n_\rmg = n + \omega \, \partial_{\omega} n$, one can assume $n_\rmg \approx n$, for low frequencies or if $n$ is independent of the angular frequency as in non-dispersive media.
Consequently, for photons, there is an $n^2$ ratio between both momenta, while defining $n = \sqrt{\epsilon_\rmr \mu_\rmr}$, and $v_\phi = (\epsilon \mu)^{-1/2}$ (for homogeneous, isotropic, non-dispersive, space-time independent media), leads to the same $n^2$ ratio for classical fields.
In dispersive media, an $n n_{\rmg}$ ratio is admitted for photons \cite{gar04} and sometimes inferred \cite{jon78,pfe07,phi11,dod12} for electromagnetic fields though no broad consensus has emerged.}
\begin{tabular}{ccc}
\hline
\hline 
& Momentum of & Momentum density \\
& a single photon\footnote{Some authors set $k = \omega / c$ with $p_{\rm Mink} = n \hbar k$, $p_{\rm Abra} = \hbar k / n_\rmg$.} & of classical fields \\
\hline \\
\textbf{Minkowski's} & $p_{\rm Mink} = \hbar k = n \hbar \dfrac{\omega}{c}$ & ${\bf g}_{{\rm Mink}} = \bfD \times \bfB$ \\
\textbf{Abraham's} & $p_{\rm Abra} = \dfrac{\hbar k}{n n_\rmg} = \dfrac{\hbar}{n_\rmg} \dfrac{\omega}{c}$ &  ${\bf g}_{{\rm Abra}} = \dfrac{\bfE \times \bfH}{c^2}$ \\
\\
\hline
\hline
\end{tabular}
\end{table}

At first glance, the dilemma seems to arise when light simply propagates through a medium, 
suggesting that the effect originates from the permittivity and/or the magnetic permeability. 
However, it does not.
The key mechanism in the Abraham-Minkowski dilemma is the modification of the wave/photon velocity compared with the speed of light in vacuum, or equivalently the modification of the wavenumber, due to various causes.
While the dilemma was only investigated in matter, this letter presents two systems with starkly different causes for light speed modification. In both cases, we obtain Abraham's and Minkowski's momenta.
The first system considers electromagnetic fields propagating inside the dispersive metallic waveguide of a vacuum electron tube.
The second system considers Langmuir waves within the Landau damping (or amplification) context in plasmas.
Our main goal is to demonstrate the universality of the Abraham-Minkowski dilemma beyond materials, so we give priority to the first system since forces and energies in vacuum are well established.
In addition, we have simulations to assess the extention of the dilemma to waveguides with an algorithm benchmarked against experimental measurements (see below).
Evidences that the dilemma occurs in plasmas (using an analogous treatment) are raised at the end of this letter.
The resolution of the dilemma for both systems goes beyond the resolution for materials since it involves a momentum flux from Maxwell's electromagnetic stress. 

Before proceeding, we should stress that, for both plasmas or waveguides, 
we describe the wave-particle interaction with the less popular $N$-body (many body or molecular dynamics) description. 
Compared to kinetic (\textit{e.g.} Vlasovian) description, it involves a huge number of degrees of freedom that seem difficult to compute, but this approach was proven feasible \cite{els03,esc18,min18} for both cases we investigate, and provides easily the two momenta of light.

Regardless of the cause, the light velocity modification leads to a refractive index $n = c / v_\phi$, with the phase velocity $v_\phi$ and, for dispersive systems ($n$ depending on the frequency), a group index $n_\rmg = c / v_\rmg$, with the group velocity $v_\rmg$.
In matter, both indices (and velocities) are naturally linked to the permittivity and the magnetic permeability.
In vacuum waveguides, like resonant cavities of slow-wave structures, it is the device geometry that determines both velocities.
Indeed, their dispersion relations can be calculated \cite{jac99} from sourceless Maxwell equations with the metallic wall boundary conditions.
For instance, the helical slow-wave structure of a traveling-wave tube \cite{gil94} (a typical electron tube used for wave amplification, similar to free-electron lasers) can slow down the phase velocity by  a factor $1/7$.
Large refractive indices (\textit{e.g.} $n \approx 7$) are frequent in waveguides, 
making the difference between Abraham's and Minkowski's momenta more dramatic.
The effects of this deceleration on wave propagation are the same as in matter, 
but it is easier to investigate the dilemma in waveguides since we can then start from Maxwell's equations in vacuum with sources.
To complete the system, we consider an electron beam propagating through the waveguide in resonance with the waves ($|\bfv| \simeq v_{\phi}$). For a moderate beam intensity, the particles do not modify the dispersion relation of the waveguide and the momentum exchange does not change the refractive index.

Deriving Abraham's and Minkowski's expressions in a waveguide requires a proper definition of the refractive and group indices in the device. 
This is done using a decomposition \cite{kuz80,and13,min19b} separating dependence on space ($\bfr$) and time ($t$) for the electric $\bfE$ and magnetic $\bfH$ fields,
\begin{align}
\bfE(\bfr,t) &= \sum_k \mathcal{V}_k (t) \bcE_k (\bfr) \, , \label{e:FieldE} \\
\bfH(\bfr,t) &= \sum_k \rmi \mathcal{I}_k (t) \bcH_k (\bfr) \, , \label{e:FieldH}
\end{align}
where the fields (with wavenumber $k$) obey the Maxwell equations in vacuum with sources. The imaginary $\rmi$ factor in Eq.~\eqref{e:FieldH} simplifies the connection below between the evolution equations obtained from Maxwell and Hamilton formalism.
A sum instead of an integral over all $k$'s is chosen for convenience. 
More accurately, a sum over all possible propagation bands is necessary, 
but, for the sake of simplicity, only the dominant mode is considered.
Using the Weyl gauge, the electric field is $\bfE = - \dot \bfA$, 
with the electromagnetic vector potential $\bfA = \sum_k \rmi \mathcal{I}_k \bcA_k$.
Since we deal with wave propagation in vacuum, the same coefficients $\mathcal{I}_k$ as for the magnetic field are used, 
though this would no longer hold in matter.

Using this field decomposition, Helmholtz' equations (sourceless Maxwell equations) read $\rot \bcE_k = - \rmi \mu_0 \omega_k \bcH_k$, and $\rot \bcH_k = \rmi \epsilon_0 \omega_k \bcE_k$, with eigenfrequencies $\omega_k$.
We normalize eigenfields $\bcE_k, \bcH_k,$ by
\begin{align}
\Omega_k &= \int_V \epsilon_0 \, | \bcE_k(\bfr) |^2 \, \rmd V  = \int_V \mu_0  \, | \bcH_k(\bfr) |^2 \, \rmd V \, . \label{e:NormalN}
\end{align}
With this decomposition, the Maxwell-Amp{\`e}re and Maxwell-Faraday evolution equations read
\begin{align} 
 \dot \cV_k (t) & = \partial_t {\cV_k} = - \cI_k (t)  \, \omega_k \,  + \sum_\imath e  \, \dot \bfq_\imath \cdot \frac{\bcE^*_k(\bfq_\imath)}{\Omega_k} \, , \label{e:evolMax1}  \\
 \dot \cI_k (t) &=  \partial_t {\cI_k} = \cV_k (t)  \,  \omega_k  \, , \label{e:evolMax2} 
\end{align}
respectively.
We then re-express the well-known electromagnetic energy \cite{jac99} (Hamiltonian of fields) in vacuum $\frac{1}{2} \int_V (\epsilon_0 |\bfE|^2 + \mu_0 |\bfH|^2) \, \rmd V$ as \cite{and13}
\begin{equation}
H_{\rm em} =  \sum_k \frac{1}{2} \Big( \cV_k (t) \cV^*_{k} (t)  +  \cI_k (t) \cI^*_{k} (t) \Big) \Omega_k  \, .
\end{equation}
In this representation, $\mathcal{V}_k$ and $\mathcal{I}_k$ are canonical variables, with $\mathcal{V}_k$ the ``conjugate momenta'' and $\mathcal{I}_k$  the ``generalized coordinates''. 
The normalisation \eqref{e:NormalN} ensures their Poisson brackets are canonical.

The dynamics of charged particles ($i$) in electromagnetic fields is dictated by the Lorentz force $\sum _i m \ddot \bfq_i = - e \dot \bfA(\bfq_i) - e (\dot \bfq_i \cdot \nabla) \bfA (\bfq_i) + \sum _i e \dot \bfq_i \times \mu_0 \bfH$,
leading to the well-known particles energy \cite{jac99,lan84} (\textit{a.k.a.} coupling Hamiltonian) $H_{\rm cpl} = \sum_i (2 m)^{-1} |\bfp_{{\rm can},i} - e \bfA(\bfq_i)|^2 $.
Relativistic expressions leave unchanged the field momentum.
From the total energy (self-consistent Hamiltonian) $H_{\rm tot} = H_{\rm em} + H_{\rm cpl}$, the Hamilton evolution equations are relations \eqref{e:evolMax1}-\eqref{e:evolMax2} if $- \rmi \omega_k \bcA_k =\bcE^*_k$, in agreement with the Weyl gauge, and on setting $\Omega_k = \omega_k$.

The Lagrangian of the system is obtained with the Legendre transformation of the total Hamiltonian, as $L_{\rm tot} = \sum_i \bfp_{{\rm can},i} \cdot \dot \bfq_i + \sum_k \cV_k \dot \cI_k - H_{\rm tot}$. This Lagrangian is invariant under space translations $\bfq_i \rightarrow \bfq_i + \delta_\bfr$ and
$\mathcal{I}_k \rightarrow \mathcal{I}_k \, \rme^{- \rmi \mathbf{k} \cdot \delta_\bfr}$.
Hence, the infinitesimal variation of the Lagrangian (\textit{a.k.a.} the generalised force on the system)
\begin{equation}
  \derivep{L_{\rm tot}}{\delta_\bfr} 
  = \deriv{\bfp_{\rm tot}}{t} 
  = \sum_i \dot \bfp_{{\rm can},i} + \sum_k \rmi \mathbf{k} \deriv{}{t} \Big( \cV^*_k \cI_k  \Big) \, ,
  \label{e:forceTOT}
\end{equation}
must vanish.
Applying Noether's theorem \cite{min18} leads to a first formulation of the conserved total momentum in the waveguide 
\begin{equation}
\bfp_{\rm tot} = \sum_i \bfp_{{\rm can},i} + \sum_k \rmi \mathcal{V}^*_k  \mathcal{I}_k \mathbf{k} \, , \label{e:TotmomNoether}
\end{equation}
as the sum of canonical momenta of the particles and the fields, with Minkowski's momentum in the waveguide
\begin{equation}
{\bf p}^{\rm wg}_{{\rm Mink}} = \sum_k \rmi \mathcal{V}^*_k  \mathcal{I}_k \mathbf{k}\, . \label{e:MinkMomentum}
\end{equation}

In addition, the derivative of Eq.~\eqref{e:NormalN} enables the calculation of the group velocity $v_\rmg =\partial_k \omega_k$, with $\omega_k = \Omega_k$. 
Using the derivative of Helmholtz' equations, the identity $\mathfrak{B} \cdot (\nabla \times \mathfrak{A}) = \nabla \cdot (\mathfrak{A} \times \mathfrak{B}) + \mathfrak{A} \cdot (\nabla \times \mathfrak{B})$, for arbitrary $\mathfrak{A}, \mathfrak{B}$, and the divergence theorem, we obtain ${\bf v}_\rmg  = (1 /\omega_k ) \int_V  \bcE^*_k(\bfr)  \times  \bcH_k(\bfr)  \, \rmd V + {\rm c.c.}$,
with ${\rm c.c.}$ denoting complex conjugate.
A simple rearrangement, with $\omega_k = k v_\phi = kc / n$ and $v_\rmg = c / n_\rmg$, leads to the relation for the wavevector
\begin{equation} 
\mathbf{k}  = \frac{n n_\rmg }{c^2} \int_V  \bcE^*_k(\bfr)  \times  \bcH_k(\bfr)  \, \rmd V + {\rm c.c.}\, , \label{e:wavenumber}
\end{equation}
in the monochromatic (\textit{a.k.a.} continuous waveform, CW) regime where group and refractive indices are considered as constant.
Then, Eq.~\eqref{e:MinkMomentum} yields the Minkowski momentum density
\begin{equation} \label{e:MinkWG}
{\bf g}^{\rm wg}_{{\rm Mink}} (\bfr,t) = \frac{n n_{\rmg}}{c^{2}} \, \bfE \times  \bfH \, ,
\end{equation}
for dispersive waveguides in the monochromatic regime (see the difference with materials in Table~\ref{t:table1}).

On the other hand, Abraham's momentum is found exactly as in vacuum, using the classic reformulation \cite{jac99} of the Lorentz force density 
\begin{equation} 
{\bf f} =  \nabla \cdot \boldsymbol{\sigma} - \partial_t (\bfE \times \bfH) / c^2 \, , \label{e:lorentzdensity}
\end{equation}
obtained from Maxwell equations in vacuum with sources, with the Maxwell stress tensor $\boldsymbol{\sigma}$, and the Abraham momentum density 
\begin{equation}
{\bf g}^{\rm wg}_{{\rm Abra}}(\bfr,t) = \frac{1}{c^2} \bfE \times  \bfH \, , \label{e:gAbra}
\end{equation}
as for materials (see Table~\ref{t:table1}).
For waveguides, we use decomposition \eqref{e:FieldE}-\eqref{e:FieldH} \cite{min18}  and Eq.~\eqref{e:wavenumber} for the wavevector to re-express Eq.~\eqref{e:gAbra} as the Abraham momentum 
\begin{equation}
{\bf p}^{\rm wg}_{{\rm Abra}} = \sum_k  \frac{1}{n n_\rmg} \rmi \mathcal{V}^*_k \mathcal{I}_k \mathbf{k} \, . \label{e:AbraMomentum}
\end{equation}

The last item needed to resolve the Abraham-Minkowski dilemma for waveguides is the link between both momenta, Eqs~\eqref{e:MinkMomentum} and \eqref{e:AbraMomentum}.
This is done by comparing the two balances of forces of the system: a first one from the Noether approach Eq.~\eqref{e:forceTOT} $\sum_i \dot \bfp_{{\rm can},i} + \dot \bfp_{\rm Mink} =0$, and a second one from the Lorentz force Eq.~\eqref{e:lorentzdensity} $\sum_i \dot \bfp_{{\rm kin},i} + \dot \bfp_{\rm Abra} -\int_V \nabla \cdot \boldsymbol{\sigma} \, \rmd V = 0$.
Comparing both leads to a second formulation for the conserved total momentum
\begin{equation}
 \bfp_{\rm tot} = \sum_i \bfp_{{\rm can},i}+ \bfp_{{\rm Mink}} = \sum_i \bfp_{{\rm kin},i}  +  \bfp_{\rm Abra}  + \bfp_{\rm Flux} \, , \label{e:totmom}
\end{equation}
with the sum running on particles $(i)$, which is the crucial relation needed to solve the dilemma.
In our field representation, this gives the momentum flux
\begin{equation}
\bfp_{{\rm Flux}} =  \sum_k \Big( 1- \frac{1}{n n_\rmg}  \Big)  \rmi \mathcal{V}^*_k \mathcal{I}_k  \mathbf{k} + \sum_i \sum_k  e \rmi  \cI_k \bcA_k(\bfq_i) \, ,
\label{eq:meth:pflux}
\end{equation} 
as the difference ensuring Eq.~\eqref{e:totmom}.

Most importantly, 
the use of the Lorentz force introduces the time derivative $\dot \bfp_{\rm Flux} = - \int_V \nabla \cdot \boldsymbol{\sigma} \, \rmd V$ of the momentum flux leaving volume $V$.
As soon as amplification or attenuation occurs, the system is not spatially homogeneous since the fields exert a stress and thereby the force balance needs to be completed by the Maxwell stress.
Indeed the time derivative of Eq.~\eqref{eq:meth:pflux} coincides with the integral of the divergence of Maxwell's stress tensor \cite{jac99} $\nabla \cdot \boldsymbol{\sigma} = \epsilon_0 (\nabla \cdot \bfE) \bfE + \mu_o (\nabla \times \bfH) \times \bfH +  \epsilon_0 (\nabla \times \bfE) \times \bfE$.

\begin{figure}
\centering
\includegraphics[width=\columnwidth]{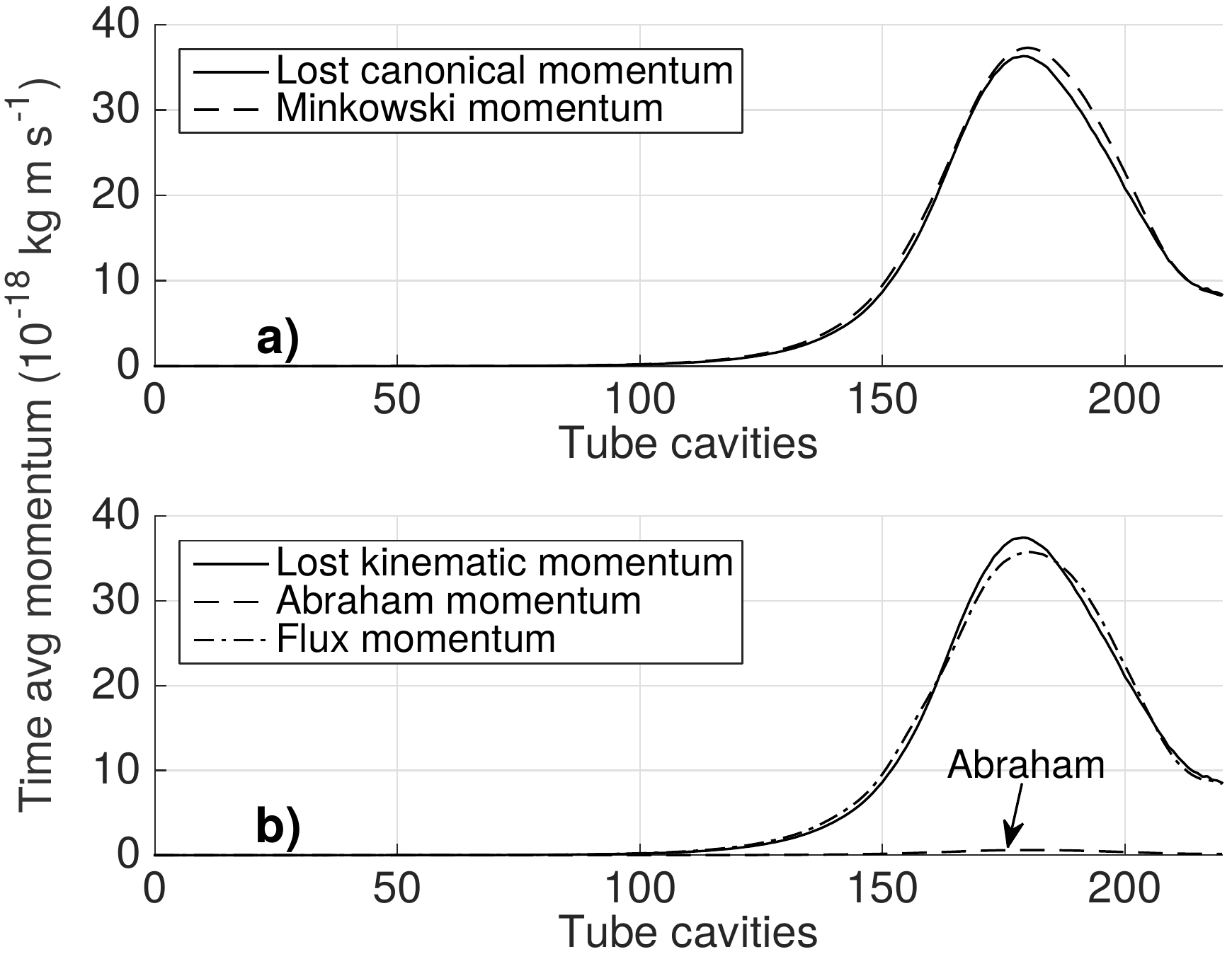}
\caption{\label{f:dimo} 
Momenta balance (time averaged per period $2 \pi / \omega_k$) per cavity (viz.\ waveguide cell) of a traveling-wave tube, 
in monochromatic regime \cite{min18}, from an algorithm \cite{and15,min17,min19} used to design industrial traveling-wave tubes.
To ensure conservation of the total momentum, we adapted the simulation to consider only the wave-particle exchange in the system and to discard any losses or external interactions.
The electron velocity is set slightly faster than the phase velocity (given by the structure geometry) to compel the wave amplification from the beginning. After 175 cavities, nonlinear trapping \cite{els03} of particles provokes the decline of the wave momentum.
a) Conservation of $\sum_i \bfp_{{\rm can},i}+ \bfp_{{\rm Mink}}$ from Eq.~\eqref{e:totmom} per cavity. Solid line: canonical momentum lost by particles (plotted positively). Dashed curve: Minkowski's momentum Eq.~\eqref{e:MinkMomentum}.
b) Conservation of $\sum_i \bfp_{{\rm kin},i}  +  \bfp_{\rm Abra}  + \bfp_{\rm Flux}$ from Eq.~\eqref{e:totmom} per cavity. Solid line: kinematic momentum lost by particles (plotted positively). 
Dashed curve: Abraham's momentum Eq.~\eqref{e:AbraMomentum}. 
Dash-dot curve: momentum flux Eq.~\eqref{eq:meth:pflux} leaving cavities. 
The conservation of total momentum \eqref{e:totmom} is observed as the particle momentum loss (plotted positively) coincides with the wave momentum increase.
The ratio between Minkowski's and Abraham's momenta is $n n_{\rmg} = 62.2$, with $n = 7.2$ and $n_{\rmg} = 8.6$, in agreement with the model parameters,
through the whole device.}
\end{figure}

The key difference of Eq.~\eqref{e:totmom} with the resolution of the dilemma in materials \cite{bar10,bar10b} is the inclusion of the momentum flux which becomes crucial for waveguide-based amplifiers (while it vanishes for mere propagation in a passive medium). 
Figure~\ref{f:dimo} shows an example of momentum exchange in a traveling-wave tube with a large $n n_{\rmg}$. 
The algorithm used is based on an $N$-body approach \cite{and13,min18} with a decomposition similar to Eqs~\eqref{e:FieldE}-\eqref{e:FieldH}, specialized for the traveling-wave tube geometry and combined with a degrees-of-freedom reduction. 
It has shown an excellent agreement with real tube measurements \cite{and15,min17,min19}. 
For Figure \ref{f:dimo}, we removed specific tube features (waveguide losses, industrial adjustments, imperfect adaptations) 
to limit the interaction only to the momentum exchange \eqref{e:totmom}.
In helix traveling-wave tubes, fields can be assumed almost longitudinal, leading to only a small difference between canonical and kinematic momenta.
Since Abraham's momentum is much smaller than Minkowski's, and because, in typical electron tubes, the difference between canonical and kinematic momenta of electrons can be really small, we immediately see that the momentum flux can obviously not be ignored.

Note that, with these new forms for Abraham's and Minkowski's expressions, we can oversimplify the quantization by enforcing $\rmi \mathcal{V}^*_k \mathcal{I}_k \rightarrow \hat{N}_k \hbar$, in Eq.~\eqref{e:MinkMomentum} and Eq.~\eqref{e:AbraMomentum}, 
to quantize classical fields in terms of photon counting operators $\hat{N}_k$. 
Though crude, this procedure reflects on the debate on photon momentum. Using this transform, Minkowski's momentum, respectively Abraham's, for a single photon 
becomes the familiar $p_{\rm Mink}  = \hbar k$, respectively $p_{\rm Abra} = \hbar k / (n n_\rmg)$ (see Table~\ref{t:table1}).

The dilemma does not only extend to vacuum waveguides 
but is relevant to every wave-particle system (with $v_\phi \neq c$) involving momentum exchange. 
Arguably the most famous one is Landau damping \cite{lan46,mou11} 
(and Landau growth, \textit{a.k.a.} bump-on-tail instability, with the same origin \cite{esc18}) 
occurring in plasma physics (beam-plasma system).
Here, the propagation medium of the Langmuir waves is a plasma 
and the dispersion relation derives from the (bulk) electron velocity and position distribution functions.
The interaction between waves and (beam) electrons is the key to Landau damping.
To investigate the latter, the $N$-body Hamiltonian approach has shown \cite{myn78,els03} 
that the associated dynamics conserves a total wave-particle momentum.
The conservation of this total momentum implies a non-linear synchronization between the particles and the wave, 
leading to the physical result that Landau damping or growth is linked 
to the slope of the particle velocity distribution function $f (\bfv_i)$ near the phase velocity of the wave.
Although it may seem surprising to relate Landau damping and vacuum waveguides,
this connection was already fruitfully made decades ago \cite{tsu,tsu91}.
Beside the transition to chaos \cite{dov05a}, Landau damping 
(more precisely the nonlinear synchronisation between an electron beam and waves) 
has been accurately investigated \cite{dov05b} using a traveling-wave tube.  
In both cases, the physics is globally the same, based on momentum exchange, 
and only differs in the origin of the dispersion relation for the wave propagation. 

Because the wave-particle interactions for traveling-wave tubes and for Landau damping are analogous \cite{dov05b,els03}, a similar spectral decomposition to Eqs~\eqref{e:FieldE}-\eqref{e:FieldH}, applies to Langmuir waves in plasma.
However, there are some critical differences. 
In particular, Langmuir waves can be considered as electrostatic, 
which implies equality of the canonical and kinematic momenta of particles ($\bfp_{{\rm can},i} = \bfp_{{\rm kin},i}$).
Yet, the Abraham-Minkowski dilemma still applies, thanks to the crucial momentum flux: 
A complete demonstration that Langmuir waves have two momenta will be the subject of a forthcoming publication.
Indeed, conservation of the total momentum, as in Eq.~\eqref{e:TotmomNoether}, has already been demonstrated experimentally \cite{dov05b} and theoretically \cite{els03,esc18} (without identifying Minkowski). 
On the other hand, when expressing the electric force $-e \nabla \varphi$, with the electrostatic potential $\varphi$ (proportional to the square of the wave intensity), 
Abraham's momentum arises complemented with the momentum flux to ensure momentum balance as well, in agreement with Eq.~\eqref{e:totmom}.

In many ways, the Abraham-Minkowski controversy looks like a modern version, involving electrodynamics, 
of the \textit{vis viva} controversy \cite{ilt71} (during the 17\textsuperscript{th} and 18\textsuperscript{th} centuries, 
both ``linear momentum'' ($mv$) and ``kinetic energy'' ($mv^2$)  were discussed as the same concept of ``force'' as it was called). 
The problem originates from the difference between definitions one uses. 

For either physical system, one obtains two different equations for the momentum conservation: 
one from the Hamilton properties leading to Minkowski's (canonical) expression, 
and one from the force balance leading to Abraham's (kinematic).
This leads to the conclusion that the Abraham-Minkowski dilemma arises in vacuum waveguides of electron tubes and in Landau damping/amplification.

Finding two momenta of light not only in dielectrics and magnetic materials, but also in waveguides and in plasmas,
implies that the difference between Abraham's and Minkowski's is a universal issue, relevant to many chapters of physics.
Investigating momentum exchange in wave-particle systems requires taking notice of the dilemma, 
because it results from the same fundamental processes:
we expect the debate to arise in every domains where momentum exchange occurs with slowed down light.
It is worth noting that simulations presented in Figure~\ref{f:dimo} were made 
with an algorithm \cite{and15,min17,min19} used for industrial applications. Consequently, the distinction between Abraham's and Minkowski's approaches
and considering the Maxwell stress, is relevant to actual devices.

Unlike for materials, solving the Abraham-Minkowski dilemma for waveguides and plasmas implies taking account of the momentum flux (which cannot be ascribed to a supporting material medium), meaning this dilemma also highlights a difference about 
the impact of the electromagnetic stress on the expression of the light momenta.

The authors gratefully thank Olivier Agullo, Didier B{\'e}nisti, Yann Camenen, Caroline Champenois, L{\'e}na{\"i}c Cou{\"e}del, Nicolas Dubuit, Dominique F.~Escande and David Zarzoso for their critical reading of the manuscript, and F.~Javier Artola, Thomas Durt, Andr{\'e} Nicolet, Valentin Pigeon, Alexandre Poy{\'e}, Brian Stout, G{\'e}rard Tayeb and Fr{\'e}d{\'e}ric Zolla for fruitful discussions.

\end{document}